\documentclass[12pt]{article}
\usepackage[dvips]{graphicx}
\usepackage{amssymb}
\usepackage{amsmath}

\def\beq{\begin{equation}}\def\eeq{\end{equation}}
\def\bea{\begin{eqnarray}}\def\eea{\end{eqnarray}}

\begin{document}
 
\title{Pilot Wave model that includes creation and annihilation of particles}
 
\author{Roman Sverdlov
\\Raman Research Institute,
\\C.V. Raman Avenue, Sadashivanagar, Bangalore -- 560080, India}

\date{November 17, 2010}
\maketitle
 
\begin{abstract}

\noindent The purpose of this paper is to come up with a Pilot Wave model of quantum field theory that incorporates particle creation and annihilation without sacrificing determinism. This has been previously attempted in \cite{visibility}, in a much less satisfactory way. In this paper I would like to "clean up" some of the things. In particular, I would like to get rid of a very unnatural concept of "visibility" of particles, which makes the model much simpler. On the other hand, I would like to add a mechanism for decoherence, which was absent in the previous version. 

\end{abstract}
 
\subsection*{1. Introduction}

As it is well known, when we consider many body quantum mechanics, we define $\psi$ function in a configuration space, rather than ordinary one. According to Everett's view, once the number of particles (or, equivalently, the dimensionality of a configuration space) is sufficiently large, the wave function splits into several non-overlapping branches. Their lack of futher interaction make them appear as "parallel universes" (even though in reality they represent different parts of the wave function in "the same" universe). The subsequent behavior within each "universe" is consistent with a collapse of wave function, even though no true collapse has occured. 

According to de Broglie and Bohm's view, there is a particle in a configuration space (or, equivalently, a set of particles in ordinary one) that co-exist with a wave function. While the wave obeys Schrodinger's equation (without any influence comming from the particle), the particle obeys \emph{guidence equation} 
\beq \vec{v} (t, \vec{x}(t)) = \frac{1}{m} \; Im \; \vec{\nabla} \; ln \; \psi \eeq
While its trajectory is clearly deterministic, it is possible to define a probability $\rho$ in a \emph{classical} sense, as a measure of "ignorance of the observer", which is, in contrast to $\psi$, is subjective. In light of the fact that $Im \vec{\nabla} \psi$ coincides with probability current, the "classical" probability $\rho$ coincides with $\vert \psi \vert^2$, as desired.  At the same time, however, if the branches split, the particle is forced to go into only one of these branches. Thus, the probability of finding it within any other branch will be nearly zero, despte the fact that the value of $\psi$ is not. This, again, leads to the appearance of "collapse" of wave function, without the one trully taking place. In the context of Pilot Wave models, this feature is called "effective collapse". 

The purpose of this paper is to redo Pilot Wave model for the case of creation and annihilation of particles. It is easy to see, however, that creation and annihilation of particles, being "discrete" can not be part of a differential equation that defines determinism (such as "guidence equation" shown above). D\"urr and others have proposed that deterministic evolution of quantum states is being "interrupted" with discrete jumps; the timing of the latter is determined by probabilistic laws (see \cite{jumps}). 

 In this paper, however, I will propose a competing view which allows the determinism to hold \emph{at all times}. I will accomplish this by claiming that there is no "true" creation/annihilation of particles. The number of particles is always fixed, but some of the particles "hide" from our view. In particular, I will extend our space $\mathbb{R}^3$ (or spacetime $\mathbb{R}^4$) by adding extra compactified coordinate $x_4 \in \Gamma$ ($x_4 + 2 \pi = x_4$), thus obtaining \emph{four} dimensional space $\mathbb{R}^3 \times \Gamma$ (or \emph{five}-dimensional spacetime $\mathbb{R}^4 \times \Gamma$). Whenever $0 \leq x_4 < \pi$ the particle is "visible" and when $\pi \leq x_4 < 2 \pi$ it is not. 

In light of the fact that the compactified $x_4$ coordinate is continuous, there are no "jumps". Thus, we are able to come up with deterministic Pilot Wave model. At the same time, during the "continuous" trajectory of a particle, a "fictitious" jump occurs when it "happens" to cross either $x_4 =0$ or $x_4 = \pi$ line. These jumps, of course, amount to creation and annihilation of particles. Their timing, however, is completely determined, since it follows from the analysis of the behavior of $x_4$ component of \emph{deterministic} trajectory of a particle.  The goal of the rest of the paper is to come up with specific Pilot Wave model that meets the above description. 

\subsection*{2. Creation and annihilation of particles}

Let us now formulate more precisely what we have said in the end of the previous section. We will introduce a notation: $\vec{x}^{(4)} = (x_1, x_2, x_3, x_4)$, and $\vec{x}^{(3)} = (x_1, x_2, x_3)$ (thus, $\vec{x}^{(4)} = (\vec{x}^{(3)}, x_4)$). In this notation, the phenomenon of particles "hiding" from our view can be expressed in definition of field operator, $\phi (\vec{x}^{(4)}= \phi(\vec{x}^{(3)}, x_4)$ as 
\beq \phi (\vec{x}^{(3)}, x_4) = \phi (\vec{x}^{(3)}), \; 0 \leq x_4 < \pi; \phi (\vec{x}^{(3)}, x_4) =1, \; \pi \leq x_4 < 2 \pi \eeq
This means that if we act with this operator on an empty state, we get
\beq \phi (\vec{x}^{(3)}, x_4) \vert 0 \rangle = \vert \vec{x}^{(3)}) \rangle, \; 0 \leq x_4 < \pi; \phi (\vec{x}^{(3)}, x_4) = \vert 0 \rangle, \; \pi \leq x_4 < 2 \pi \eeq
Now, in terms of $\vec{x}^{(4)}$ the left hand side of the above equations always amounts to ONE particle being present, regardless of the vaues of $x_4$. The right hand side, on the other hand, can give us \emph{either} one particle state (if $0 \leq x_4 < \pi$) \emph{or} empty state (if $\pi \leq x_4 < 2 \pi$). This essentially defines what we meant when we say particles "hide". 

For simplicity we will assume from now on that there is only one kind of particles, and it happens to be spin $0$. Then the configuration space is $\mathbb{R}^{4N}$, where $N$ is the total number of particles (or, in other words, the \emph{largest} possible number of "visible" ones). Then we can define the probability amplitude in $\mathbb{R}^{4N}$ as 
\beq \psi (x_1^{(4)}, . . . , x_N^{(4)}) = \langle 0 \vert \phi (\vec{x}_1^{(4)}) . . . \phi (\vec{x}_N^{(4)}) e^{iH (t - t_0)} \vert \psi (t_0) \rangle \eeq
It shoud be emphasized that on the right hand side of the above equation, both $\vert \psi(t) \rangle = e^{iH (t-t_0)} \vert \psi (t_0) \rangle$, as well as the Hamiltonian $H$ are defined in terms of $\vec{x}^{(3)}$, \emph{not} $\vec{x}^{(4)}$. The only part where $\vec{x}^{(4)}$ comes in is the definition of field operators $\vec{x}_k^{(4)}$. The reason we can "mix" $\vec{x}^{(3)}$ with $\vec{x}^{(4)}$ is that we have an expression of $\phi (\vec{x}^{(4)})$ in terms of $\phi (\vec{x}^{(3)})$ that we mentioned earlier. 

Now, in light of the fact that we are dealing with an effective theory, $\psi$ is not normalized due to lack of unitarity. Therefore, the normalization should be included in the "translation" from $\psi$ to the "probability density", 
\beq \rho (\vec{x}_1^{(4)}, . . . , x_N^{(4)}) = \frac{ \vert \psi (\vec{x}_1^{(4)}, . . . , x_N^{(4)}) \vert^2}{\int d^4 x_1 . . . d^4 x_N \; \vert \psi (\vec{x}_1^{(4)}, . . . , x_N^{(4)}) \vert^2} \eeq
The above, however, is not the \emph{actual} probabililty density; it is only a "desired" one. We now have to come up with a \emph{deterministic} Pilot Wave model that reproduces the "desired" \emph{classical} probability. We notice that creation and annihilation of particles amounts to "transition" from one semicircle to the other. Thus, if the particle is "in the middle" of one of the semicircles, and the Hamiltonian "wants" it to make a transition, it has to "move" towards the edge of the semicircle. After all, its trajectory has to be continuus, so it can not "jump". At the same time, however, the probability amplitude \emph{within} each semicircle is constant. Thus, a particle has to "look" at the probability amplitude at the other semicircle in order to "decide" how it moves. In other words, while its motion is continuus, it depends on non-local information. 

Another suddlety that needs to be adressed is the fact that the probability amplitude is discontinuus "at the edge" of a semicircle. This, however, does not "disturb" the continuity of the trajectory of the particle. This difficulty can be adressed simulteneously with a previous one: if "most" of the information particle uses in order to "decide" how to move is non-local, then the "local" discontinuity might end up being irrelevent! In fact, electrostatics is a very good example of a theory where step function charge distribution produces continuous electric field. The key to this "paradox" is "non-local" nature of Coulumb's law. 

Apart from the fact that electrostaticcs is a good theory that avoids the above mentioned difficulties, it also happens to give us the equation that we want. After all, our goal is to define  $\vec{v} (x_1^{(4)}), . . . , \vec{v} (x_N^{(4)})$ that satisfies a continuity equation
\beq \frac{\partial \rho}{\partial t} = \vec{\nabla} \cdot (\rho \vec{v}) \eeq
Now, if we replace the above with an "electrostatic" problem with "electric field" $\vec{E}$ and "charge density" $\sigma$ where 
\beq \vec{E} = \rho \vec{v} \; ; \; \sigma = - \frac{\partial \rho}{\partial t} \eeq
then "Coulumb's law" implies a continuity equation that we are looking for \footnote{The "electrostatic" idea was also suggested in \cite{Epstein} although I don't think that they had in mind the constriction presented in this paper}! It is important to understand, however, that $\sigma$ and $\vec{E}$ do \emph{not} refer to literal charge and electric field. In fact, in our example, if $\phi$ is real scalar field (as opposed to complex) than all particles are electrically neutral. The only reason we make a reference to electrostatics is that it provides for us the mathematical structure that we are looking for.

Now, in order to be able to solve the above equation for $\vec{v}$, we have to assume that $\rho$ is known. This leads one to ask: how can $\rho$ be known if it is a \emph{classical} probability resulting from Pilot Wave model that we don't have yet? The answer to this is that $\rho$ is \emph{not} the actual probability; it is only a \emph{desired} one. Strictly speaking, $\rho$ is a filed, \emph{not} a probability density. Our \emph{goal} is to find Pilot Wave model in which it \emph{happens} to coincide with probability. 

We now use the method of images in order to compute $\vec{E}$ in compactified geometry. We consider an imaginary situation where coordinates $x_{4k}$ extend to infinity, and charge density is periodic in these coordinates, 
\beq \sigma (\vec{x}) = \sigma (\vec{x} + 2 \pi \hat{x}_{4k}) \eeq
By translational symmetry, we notice that  
\beq \vec{E} (\vec{x}) = \vec{E} (\vec{x} + 2 \pi \hat{x}_{4k}) \eeq 
from which it is easy to see that the identical copy of $\vec{E}$ satisfies the same differential equation in the compactified geometry, without having any discontinuities at $x_{4k} = 2 \pi n$. Thus, this is a solution we are seeking. Now, from the non-compact case, we can show that the area of the sphere is given by  
\beq A = \frac{2 \pi^{2N}}{(2N-1)!} r^{4N-1}\eeq
This tells us that the electric field is 
\beq \vec{E} (\vec{x})= \frac{(2N-1)!}{2 \pi^{2N}} \sum_{a_1, . . . , a_N} \int d^{4N} x' \frac{ \sigma (\vec{x}' )(\vec{x}' - \vec{x} + \vec{R}_{a_1, . . . , a_N})}{ \vert \vec{x}' - \vec{x} + \vec{R}_{a_1, . . . , a_N} \vert^{4N}} \eeq
where
\beq \vec{R}_{a_1, . . . , a_N} = \sum 2 \pi a_i \hat{x}_{4i} \eeq
is a displacement of an "image charge". Now by substituting back 
\beq \vec{E} \rightarrow \rho^{(4)} \vec{v} \; ; \; \sigma \rightarrow \frac{\partial \vert \psi \vert^2}{\partial t} \eeq
we obtain a guidance equation
\beq \vec{v}(\vec{x}) = \frac{(2N-1)!}{2 \pi^{2N} \rho^{(4)}(\vec{x})} \frac{\partial}{\partial t} \sum_{a_1, . . . , a_N} \int d^{4N} x' \; \frac{ \rho^{(4)} (\vec{x}') (\vec{x}' - \vec{x} + \vec{R}_{a_1, . . . , a_N})}{ \vert \vec{x}' - \vec{x} + \vec{R}_{a_1, . . . , a_N} \vert^{4N}} \eeq

\subsection*{3. Decoherence}

So far we have successfully came up with a model that reproduces desired probability density. However, there is one more ingredient to Pilot Wave model: if decoherence occurs, or, in other words, a wave function splits into branches, the subsequent probabiity of finding the particle within the "wrong" branches should be nearly zero (despite $\vert \psi \vert^2$ being large), and the correlation between $\rho$ and $\vert \psi \vert^2$ holds only within \emph{one} branch. 

In case of "standard" Pilot Wave model this is due to the fact that probability of finding a particle between the branches is zero. Thus, once the particle "flew" into one of the branches it is "stuck" there. The good news is that, in our case, we also have the probability density nearly zero between the branches. The bad news, however, is that the electric charge density, $\sigma = - \partial \rho / \partial t$ is nearly zero between the branches, so nothing keeps the particle from that region. This might, as suggested in the end of chapter 3 of \cite{Epstein}, lead to unwanted transitions between universes back and forth.

Since "good" and "bad" news seem to contradict each other, let us first see which is, in fact, the case. By looking at the equation $\vec{E} = \rho \vec{v}$ we can see that whenever $\rho$ is small $\vec{v}$ is large. From this we can guess that the reason that the probability of finding the particle is small is \emph{not} because a particle never gets to the $\rho \approx 0$ region but, instead, because it "flighs away very fast" \emph{once} it is there. Thus, a particle \emph{can} be between the branches \emph{as long as} it returns to \emph{one} of the branches "fast enough".

Now in the above argument it is not necessary that it should return to the same branch. It might as well return to a different one! In fact, since velocity is co-directional with "electric field", the particle might leave one branch as it follows "electric field" lines and "run into" a different branch before it ever has chance to return. We, therefore, need to modify our theory in such a way that this wouldn't happen. Instead of "gluing" a particle to a particular branch, I instead propose to introduce additional term in Hamiltonian that would "get rid" of all of the unwanted branches. In particular, I propose Hamiltonian
\beq H = H_Q + H_B \eeq
where $H_Q$ is simply borrowed from quantum field theory, while $H_B$ describes the interaction with a beable.  The $H_B$ part of Hamiltonian is designed in such a way that the wave function within any of the branches \emph{not} occupied by a beable dies out, while the wave function within the one that \emph{is} occupied is left unchanged. 

We have to be a bit careful, though. Even if the other branches will "die off" the particle will continue to periodically leave the "home branch" as it follows the electric field lines. Even though it spends very little time outside the branch, if we wait long enough the effects will accumulate so even the branch that we want to "keep" will be "destroyed". This issue can be adressed by remembering that we have \emph{already} stated that $\vert \psi \vert^2$ is not properly normalized. Our previous reason for this was the lack of unitarity of the theory. Our new reason is that $\vert \psi \vert^2$ systematically dies out due to $H_B$. But the result is still the same. Within the non-unitarity context we have adressed it by introducing normalization factor in definition of $\rho$, 
\beq \rho (\vec{x}_1^{(4)}, . . . , x_N^{(4)}) = \frac{ \vert \psi (\vec{x}_1^{(4)}, . . . , x_N^{(4)}) \vert^2}{\int d^4 x_1 . . . d^4 x_N \; \vert \psi (\vec{x}_1^{(4)}, . . . , x_N^{(4)}) \vert^2} \eeq
This same normalization adresses our "newer" concern as well. If \emph{at some point in time} a particle spends sufficiently long time within a branch number $k$, then, within this time interval, the branches $l \neq k$ will become very small. As a result, the particle will continue to spend much longer time within a branch number $k$ than it will within $l \neq k$. Thus, in proportion to their sizes, branches $l \neq k$ will coninue to die out much "faster" than the branch number $k$ does. As a result, the integral of \emph{normalized} $\rho$ over branch number $k$ will continue to be nearly $1$. 

To put it another way, $H_B$ diminishes the value of wave function in any given branch as long as a particle is \emph{not} in that branch. Thus, if it is in the gap between the branches, then \emph{all} of them "die off"; but, in light of the normalization, the "uniform" impact on $\psi$ has no bearing on $\rho$. Now, if branch number $1$ happened to be "bigger" than branch number $2$, then the former "slows down" the particle more than the latter. Thus, the time the particle spends \emph{outside} branch 1 is smaller, and, therefore, branch $1$ dies out at slower rate.  

\emph{If} we only had these two branches, then due to normalization of $\rho$, branch $1$ would be \emph{in}creasing, while branch $2$ decreasing, until the latter is non-existent. Now, if two branches are \emph{exactly} the same, then the particle might "accidentally" spend a little bit more time in one of them (simply because it happened to be closer to that branch), and as a result that one branch becomes "slightly" bigger. And, once it does, the difference grows more and more at increasing rate until the other branch is completely gone.

 In other words, the setting with more then one equal sized branch is "unstable equilibrium". The single-branch setting is semi-stable: the dynamics of a beable will \emph{not} restore other branches. But, from time to time, a branch "breaks" into few sub-branches due to decoherence. Then, all of the sub-branches except one disappear, due to the above mechanism. Thus, the measuring process includes \emph{two} steps: a decoherence \emph{and} a distraction of unwanted branches. 

Let us now go ahead and come up with a precise definition of $H_B$. In order to do this, we have to "define" the notion of being "in the same branch" as our particle. We will do that by imagining the following scenario. Our "beable" particle $B$ continuously creates $C$ particles. In order to avoid unwanted singularities, we will replace $\delta$-function with Gaussian and claim that the probability of "creation" is proportional to $e^{k \vert \vec{x} - \vec{x}_B \vert^2}$.

The $C$-particles are undergoing random walk; but, at each "step" of a random walk they can be distroyed with probability $c / \rho$. If $c$ is very small, then as long as $\rho$ is sufficiently large, the probability of disraction is virtually $0$. But, once $\rho$ becomes much smaller than $c$, the particle is almost guaranteed to be destroyed. Thus, if $c$ is chosen in such a way that $\rho \ll c$ holds true between branches but not within each branch, then $\chi \gg 0$ will be a good definition of the interior of a branch. Thus, the dynamics is
\beq \frac{\partial \chi}{\partial t} = a \nabla^2 \chi + be^{k \vert \vec{x} - \vec{x}_B \vert^2} - \frac{c \chi(\vec{x})}{\rho (\vec{x})}, \eeq
It is important to point out, however, that we \emph{do not} want to introduce \emph{literal} C-particles. After all, as explained earlier, creation and annihilation of particles is the very thing we are trying to avoid in order to restore determinism. Thus, we will now say that C particles don't exist, and $\chi$ is just a field that \emph{happened} to satisfy the above differential equation. The notion of C-particles was simply an intuitive device helped us find a differential equation that meets our purposes. 

Now, the simplest thing we can now do is to define $H_B$ as an \emph{imaginary} expression $-i \delta / \chi$, where $\delta$ is some small constant. From the intuition we have from probability theory, we know that $\chi > 0$. Thus, if $\chi$ is reasonably large, $e^{-iH_B t} \approx 1$. At the same time, if $\chi \ll \delta$, we get $e^{-iH_B t} \approx 0$ which, of course, would diminish the value of $\psi$ within the "wrong" branches. We have to be a bit careful, since $H$ was defined on ($\vec{x}^{(3)}$-based) Fock space, and \emph{not} on $\mathbb{R}^{4N}$ we are working on. 

In order to draw a link between the two spaces, for every state $\vert s \rangle$ in our Fock space we will define a \emph{set} of points $S( \vert s \rangle)$ in our configuration space that correspond to $\vert s\rangle$. From our construction we know that, as long as $x_4$ and $x'_4$ fall within the same "semicircle" (that is, either $0\leq x_4 < x'_4 < \pi$ or $\pi\leq x_4 < x'_4 < 2 \pi$) the field operators $\phi (\vec{x}^{(3)}, x_4)$ and $\phi (\vec{x}^{(3)}, x'_4)$ are the same, as long as we are using the same value of $\vec{x}^{(3)}$ in both cases. This means that, $(\vec{x}^{(3)}, x_4)$ and  $(\vec{x}^{(3)}, x'_4)$ are elements of \emph{the same} $\vert s \rangle$. We will now define $H_B$ on the Fock space as follows:
\beq H_B \vert s \rangle = - \Big( \int_{S (\vert s \rangle)} d^{4N} x \; \frac{i \delta}{\chi (\vec{x})} \Big) \vert s \rangle \eeq
In light of the fact that $H_Q$ and $H_B$ are both functions of $\vert s \rangle$, it is easy to see that the resulting value of $\psi$ will be constant within the region $S(\vert s \rangle)$. This implies that $\rho$ is constant throughout the set, as well. It is possible, however, that $\chi$ might not be constant due to the geometric shape of the branch. It is, however, reasonable to expect that $\chi$ will either be consistenly "large" or consistently "small" throughout each set, which would avoid unwanted ambiguities in $H_B$. More rigourous verification of this assertion is subject to further research. 

\subsection*{5. Conclusions}

In this paper we were able to describe creation and annihilation of particles deterministically. It was done by viewing these as a "shaddows" fo a \emph{continuous} process, namely a particle moving around a circle defined by extra compactified coordinate $x_4$ (subject to compactification $x_4 + 2 \pi = x_4$). The partilce \emph{happens} to be "visible" when it is on the side of a circle $0 \leq x_4 < \pi$, and it is "invisible" when it is on the side of a circle $\pi \leq x_4 < 2 \pi$. 

This paper should be compared to a previous one I have written which has similar ideology (\cite{visibility}). According to that other paper, a visibility is a \emph{differential} approximation to step function, $f(x_4)$. Thus, in light of continuity, $f$ has a "transition region" from $0$ to $1$. This means that a particle is neither visible nor invisible but more like "half a particle" in that region.

In this current paper I have removed $f$ entirely. The reason for this is that, from the analogy with electrostatics, I realized that even if $f$ is an exact step function, the resulting velocity field might still be continuous (just like an exact step function charge distribution produces continuous electric field). However, I plan to leave $f$ standing in the other paper, because it might add mathematical rigour to the proof of differentiability and determinism. Plus, being a believer in "intellectual pluralism" I would like to have both options "available" for future research. 

I would like to say that my idea of removing $f$ is largely a result of my communication with Struyve where he asked me to explore the modifications to the experimental predictions that $f$ might lead to. While he, himself, did not encourage me to remove $f$, my own answer to his question did. I realized that I was doing my best to taylor $f$ in such a way that the predicted modifications are as small as possible. This, naturally, leads one to ask: how is it possible that accounts for very small modifications plays such a crucial role? Then, thinking about this question lead me to see that I would still have desired differentiability of $\vec{v}$ \emph{even if} $f$ is an \emph{exact} step function! This lead me to write this new article. 

I believe that the idea, as presented in this new article, is a lot more beautiful. In previous article I was emphasizing the concept of "in-between" visibility states. It might be seen between the lines that "in between" visibility is a sign of desperation. On the other hand, in this current article, I put emphasis \emph{solely} on extra compactified dimension. I show that this dimension, alone, leads to the desired result. While this is still a bit artificial, in my opinion it is less so. 

Another major difference between the previous paper and the current one is that I added another section (section 4) in which I introduced a mechanism that allows a particle to "get rid of" unwanted "branches".  In the previous version of this theory I allowed all of the branches to exist. This posed a problem because, due to non-local nature of "Coulumb's law", a particle can potentially go from branch to branch. This situation is unique to my model because Coulumb's law, as contrasted with more standard quidence equation, is non-local.

In my previous version (\cite{visibility}) I proposed a qualitative argument as to why the transition between branches is unlikely. In particular, I proposed speculative argument to the effect that  the "electric field lines" of different branches do not intersect. But, after having talked about it with Struyve at the DICE2010 conference, he has pointed out to me some of the mistakes in the argument, and encouraged me to leave it as an open question. Being unhappy with an idea of an "open question" I decided to, instead, propose a model in which the unwanted branches disappear. Nevertheless, it should be acknowledged that the model proposed is quite unnatural, and it is crucial to be on a lookout for better models. 

In other words, according to the previous version, multiple branches exist as "parallel universes" while a particle is attached to just one of them. According to the current version, however, a particle "destroys" unwanted branches (but, at the same time, a particle is free to go from branch to branch \emph{until} they are destoryed).  

Appart from the significance this change has on my work, I belive it also introduces a new concept on a more major scale. According to most (if not all) existing Pilot Wave models, a wave has \emph{one way} influence on a particle through guidence equation. In this new model, however, the influence became \emph{two way}: a wave "guides" a particle, while a particle "destroys" the unwanted branches of a wave. 

I also believe that this new theory ade a significant revision to our definition of measurement. Up till now there are two views of measurement. According to Everett, the measurement is one-step process of a wave function "splitting" into branches. According to Bohm, it is two-step process: a wave function splits into branches \emph{and} a particle flows into one of them. The second step, however, is not really a step since a particle is \emph{already} in one of the branches at a point of the split. 

According to the proposed view, however, the measurement is \emph{three step} process: first branches split, then particle flows into one of them, and then the particle "destroys" the rest of the branches. I believe that this three-step view is also ideologically "more balanced" (while Bohm and Everett can be viewed as two opposite ideological extremes). According to Everett, the "reality" is just a wave, and there is no particle altogether. According to Bohm, the "reality" is a partice; a wave is simply guiding it. On the other hand, according to the proposed veiw, a reality consists of interplay between the two.

\end{document}